\documentclass[preprint,12pt]{elsarticle}

\usepackage{amssymb}
\usepackage{amsthm}
\usepackage{framed} 
\usepackage{amsmath,color}
\usepackage{mathrsfs}
\usepackage{graphicx}
\usepackage{epstopdf}
\usepackage{float}
\usepackage{caption}
\usepackage{subcaption}
\usepackage{bm}
\usepackage{bbm}
\usepackage{mathrsfs}
\usepackage{cleveref}
\usepackage{soul}
\usepackage{accents}
\usepackage{color,soul} 
\usepackage{color} 
\usepackage{nomencl} 
\biboptions{sort&compress}
\soulregister\citep7 
\soulregister\citet7 
\soulregister\citealp7 
\newsavebox{\measurebox} 
\usepackage{titlesec} 
\usepackage[nomarkers,figuresonly]{endfloat} 
\usepackage{tabu} 
\usepackage{longtable}
\usepackage{multirow} 

\journal{Journal of Applied Mechanics}

\makeatletter
\def\@author#1{\g@addto@macro\elsauthors{\normalsize%
    \def\baselinestretch{1}%
    \upshape\authorsep#1\unskip\textsuperscript{%
      \ifx\@fnmark\@empty\else\unskip\sep\@fnmark\let\sep=,\fi
      \ifx\@corref\@empty\else\unskip\sep\@corref\let\sep=,\fi
      }%
    \def\authorsep{\unskip,\space}%
    \global\let\@fnmark\@empty
    \global\let\@corref\@empty  
    \global\let\sep\@empty}%
    \@eadauthor={#1}
}
\makeatother

\titleformat{\paragraph}
{\normalfont\normalsize\itshape}{\theparagraph}{1em}{}
\titlespacing*{\paragraph}
{0pt}{3.25ex plus 1ex minus .2ex}{1.5ex plus .2ex}

\begin{document}

\begin{frontmatter}



\title{Crack growth resistance in metallic alloys: the role of isotropic versus kinematic hardening}


\author{Emilio Mart\'{\i}nez-Pa\~neda\corref{cor1}\fnref{Cam}}
\ead{mail@empaneda.com}

\author{Norman A. Fleck\fnref{Cam}}

\address[Cam]{Department of Engineering, Cambridge University, CB2 1PZ Cambridge, UK}

\cortext[cor1]{Corresponding author. Tel: +44 1223 748525; fax: +44 1223 332662.}

\begin{abstract}
The sensitivity of crack growth resistance to the choice of isotropic or kinematic hardening is investigated. Monotonic mode I crack advance under small scale yielding conditions is modelled via a cohesive zone formulation endowed with a traction-separation law. R-curves are computed for materials that exhibit linear or power law hardening. Kinematic hardening leads to an enhanced crack growth resistance relative to isotropic hardening. Moreover, kinematic hardening requires greater crack extension to achieve the steady state. These differences are traced to the non-proportional loading of material elements near the crack tip as the crack advances. The sensitivity of the R-curve to the cohesive zone properties and to the level of material strain hardening is explored for both isotropic and kinematic hardening.
\end{abstract}

\begin{keyword}

Kinematic hardening \sep Isotropic hardening \sep Cohesive zone modelling \sep Finite element analysis \sep Fracture



\end{keyword}

\end{frontmatter}



\section{Introduction}
\label{Sec:Introduction}

It is well established that material elements near a mode I crack tip undergo non-proportional straining due to crack advance, see for example the early analysis of crack growth by Rice and Sorensen \cite{Rice1978}. The degree of hysteresis associated with this non-proportional loading is sensitive to the nature of the hardening law of the solid. For example, it is to be expected that kinematic hardening leads to greater hysteresis than isotropic hardening. Consequently, one might expect that the choice of plastic hardening law will influence the stress intensity factor $K$ versus crack extension $\Delta a$ response, widely known as the R-curve. However, little attention has been paid to the effect of the hardening law upon crack growth resistance and no clear picture emerges from the literature. Lam and McMeeking \cite{Lam1984} analysed steady state crack tip fields and concluded from a crack opening displacement criterion that isotropic hardening augments crack growth resistance. Carpinteri \cite{Carpinteri1985} performed finite element analyses of crack propagation by means of a strain-based criterion and observed a greater amount of crack extension in the kinematic hardening case for a given remote load; this also suggests that isotropic hardening increases crack growth resistance. In contrast, we shall demonstrate that kinematic hardening significantly raises the level of plastic dissipation and, thereby, elevates the R-curve along with the steady state fracture toughness $K_{SS}$.

\section{Numerical model}
\label{Sec:NumModel}

We consider the small scale yielding problem of a plane strain mode I crack subjected to a remote stress intensity factor $K$. The elasto-plastic solid is isotropic with a Young's modulus $E$, a Poisson's ratio $\nu$ and an initial yield strength $\sigma_0$. Throughout our study we shall take $\nu=0.3$ and $\sigma_0/E=0.003$. We denote the Cauchy stress by $\sigma_{ij}$ and define $s_{ij}$ as its deviatoric part, such that $s_{ij}=\sigma_{ij} - \delta_{ij} \sigma_{kk}/3$. The plastic response involves either isotropic or kinematic hardening, as follows.\\

\noindent \textit{Isotropic hardening:} The yield condition reads
\begin{equation}
\Phi = \sigma_e - \sigma_Y = 0
\end{equation}

\noindent where $\sigma_e$ is the von Mises effective stress and the current yield strength $\sigma_Y$ is a function of the accumulated von Mises plastic strain $\varepsilon_e$. Accordingly, the effective stress $\sigma_e$ in J2 plasticity is defined as
\begin{equation}
\sigma_e^2= \frac{3}{2} s_{ij} s_{ij}
\end{equation}

The increment in plastic strain $\dot{\varepsilon}_{ij}^p$ is computed from the normality hypothesis,
\begin{equation}
\dot{\varepsilon}_{ij}^p =\dot{\varepsilon}_e \frac{\partial \Phi}{\partial \sigma_{ij}}= \dot{\varepsilon}_e \frac{3}{2} \frac{s_{ij}}{\sigma_Y}
\end{equation}

\noindent in terms of the increment in effective plastic strain $ \dot{\varepsilon}_e$. The relation between $\sigma_Y$ and $\varepsilon_e$ is given by the uniaxial tensile response, such that the true tensile stress $\sigma$ is related to the true tensile plastic strain $\varepsilon^p$ by
\begin{equation}\label{Eq:PowerLaw}
\sigma = \sigma_0 \left( 1 + \frac{E \varepsilon^p}{\sigma_0}\right)^N
\end{equation}

\noindent where $N$ is the strain hardening exponent. In addition to this power law description we also consider the case of linear hardening by taking $N=1$ and by replacing $E$ in (\ref{Eq:PowerLaw}) with the tangent modulus $E_t$.\\

\noindent \textit{Kinematic hardening}: Assume that the centre of the yield surface is located at the point $\alpha_{ij}$ in deviatoric stress space. We shall refer to $\alpha_{ij}$ as a backstress, and assume that the Armstrong-Frederick non-linear rule \cite{Frederick2007} defines the evolution of this backstress, such that
\begin{equation}\label{Eq:Frederick}
\dot{\alpha}_{ij} = c \frac{\left( s_{ij} - \alpha_{ij} \right)}{\sigma_0} \dot{\varepsilon}_e - \gamma \alpha_{ij} \dot{\varepsilon}_e
\end{equation}

\noindent where $c$ and $\gamma$ are material constants. This rule reproduces ratchetting when a material element is subjected to a non-zero mean stress and cyclic loading but predicts a particular shape of the stress-strain curve (see for example \cite{Bower2009}). In order to model a more general shape of the uniaxial tensile response the constitutive statement (\ref{Eq:Frederick}) has been extended by Chaboche \cite{Chaboche1991}. He replaced the single backstress $\alpha_{ij}$ by a finite number $n$ of backstresses $\alpha_{ij}^k$, such that
\begin{equation}\label{Eq:Chaboche1}
\alpha_{ij} = \sum_{k=1}^n \alpha_{ij}^k
\end{equation} 

\begin{figure}[H]
        \begin{subfigure}[h]{1\textwidth}
                \centering
                \includegraphics[scale=0.87]{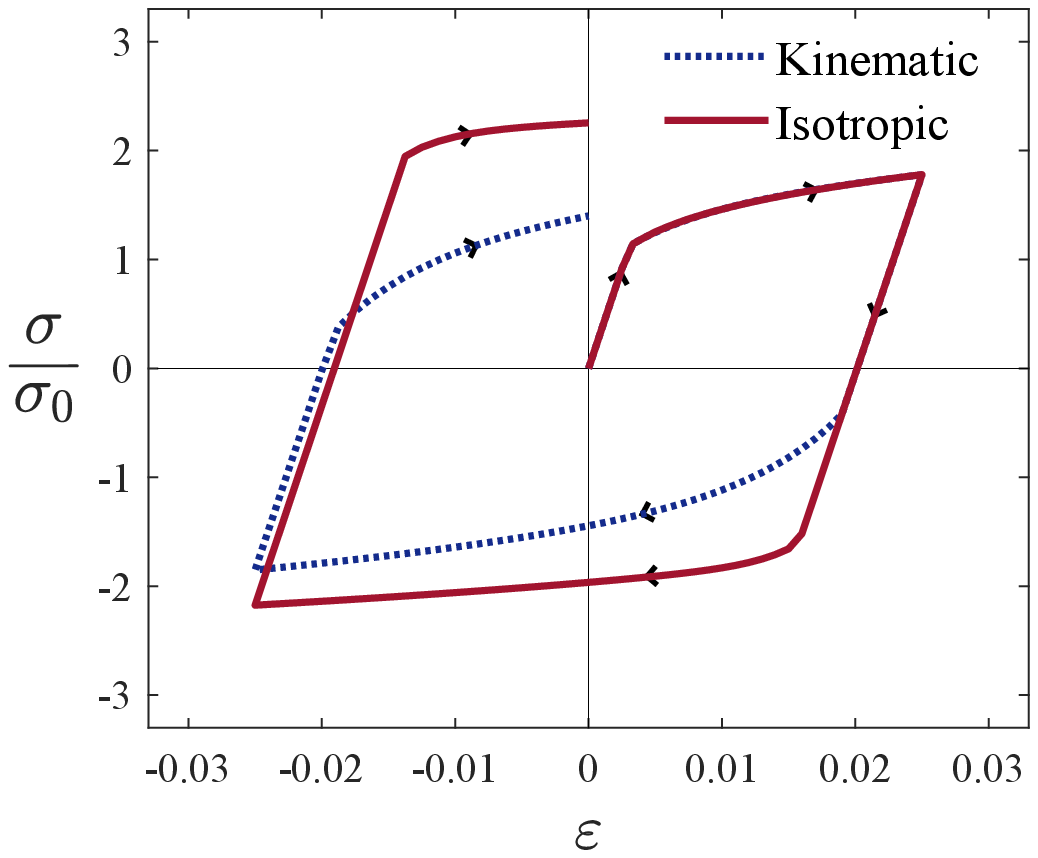}
                \caption{}
                \label{fig:Cyclic}
        \end{subfigure}

        \begin{subfigure}[h]{1\textwidth}
                \centering
                \includegraphics[scale=0.83]{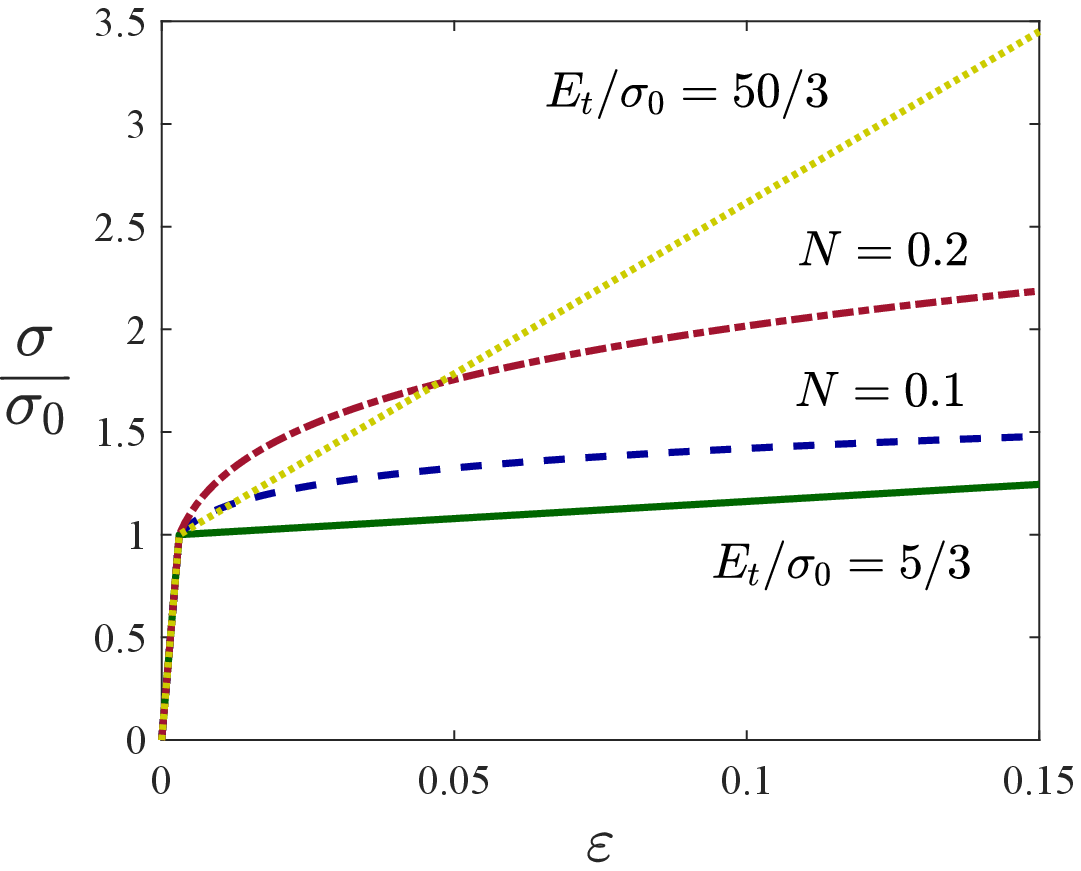}
                \caption{}
                \label{fig:Uniaxial1}
        \end{subfigure}       
        \caption{Uniaxial stress strain response for (a) cyclic loading of a non-linear hardening solid with $N=0.2$, and (b) half-cycle for linear and non-linear hardening. Material properties: $\sigma_0/E=0.003$.}\label{fig:Uniaxial}
\end{figure}

\noindent Each backstress $\alpha_{ij}^k$ evolves with $\dot{\varepsilon}_{ij}^p$ according to the independent hardening rule,
\begin{equation}\label{Eq:Chaboche2}
\dot{\alpha}_{ij}^k = c^k \frac{\left( \sigma_{ij} - \alpha_{ij} \right)}{\sigma_0} \dot{\varepsilon}_e - \gamma^k \alpha_{ij}^k \dot{\varepsilon}_e \,\,\,\,\,\,\ \textnormal{(no sum on} \,\, k \textnormal{)}
\end{equation}

\noindent in terms of the material constants $c^k$ and $\gamma^k$. The resulting Chaboche-Armstrong-Frederick (CAF) model has been widely used to capture ratchetting effects and non-linear hardening under non-proportional cyclic loading \cite{Lemaitre1990,Chaboche2008}. We consider here the case of power law hardening (\ref{Eq:PowerLaw}) and select the values of $\left( c^k, \, \gamma^k \right)$ for $k \in (1, n)$ such that the desired response in unaxial tension is obtained. The choice of $n=10$ brings the CAF model into alignment with (\ref{Eq:PowerLaw}) to within 0.04\% for the range of true tensile strain $0 \leq \varepsilon \leq 2.0$. The uniaxial stress-strain response for cyclic loading is given in Fig. \ref{fig:Cyclic} for the case $N=0.2$. In the case of linear hardening (\ref{Eq:Frederick}) is used instead of (\ref{Eq:Chaboche1})-(\ref{Eq:Chaboche2}), with $\gamma=0$ and $c=E_t$; this is the familiar Ziegler formulation \cite{Ziegler1959}. The hardening laws employed are shown in Fig. \ref{fig:Uniaxial1}; power law hardening for the choices $N=0.1$ and $N=0.2$, and linear hardening for $E_t/\sigma_0=5/3$ and $E_t/\sigma_0=50/3$.\\

\begin{figure}[H] 
    \centering
    \includegraphics[scale=0.8]{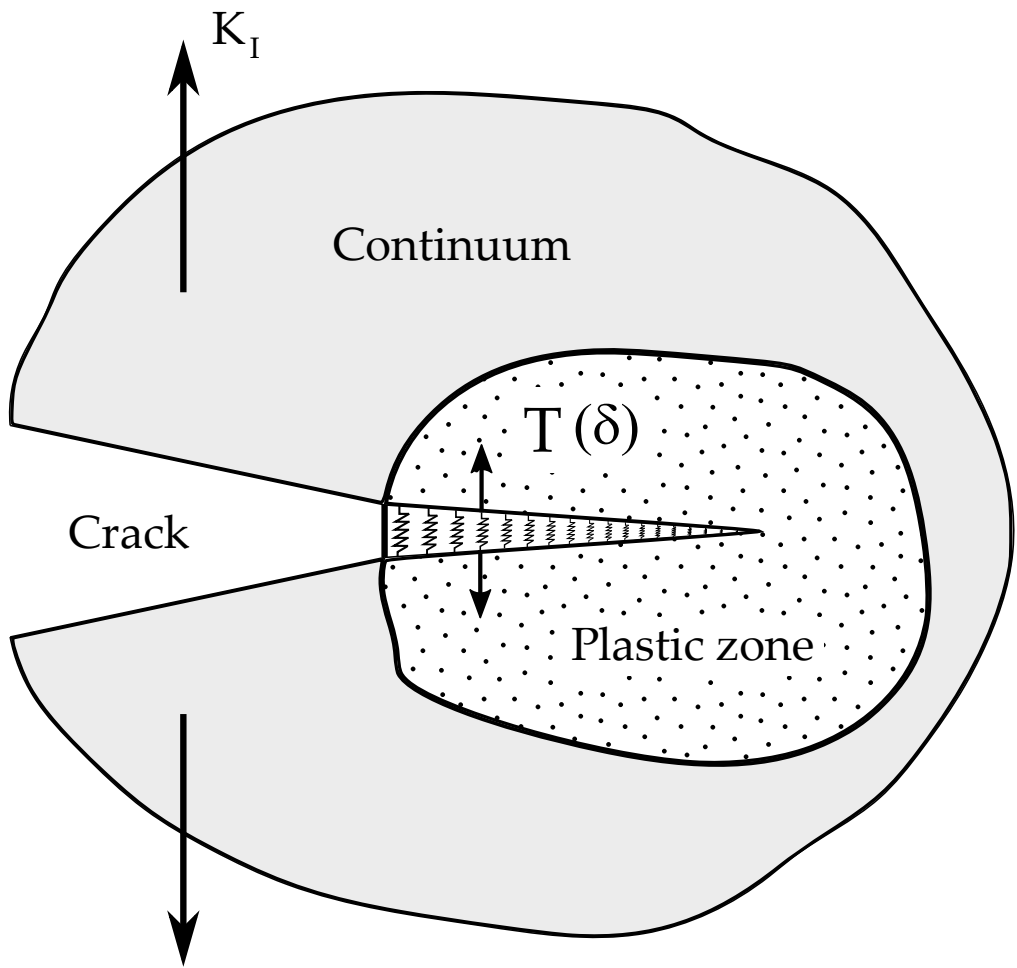}
    \caption{Schematic representation of the cohesive zone model for fracture.}
    \label{fig:Crack}
\end{figure}

\begin{figure}[H] 
    \centering
    \includegraphics[scale=0.6]{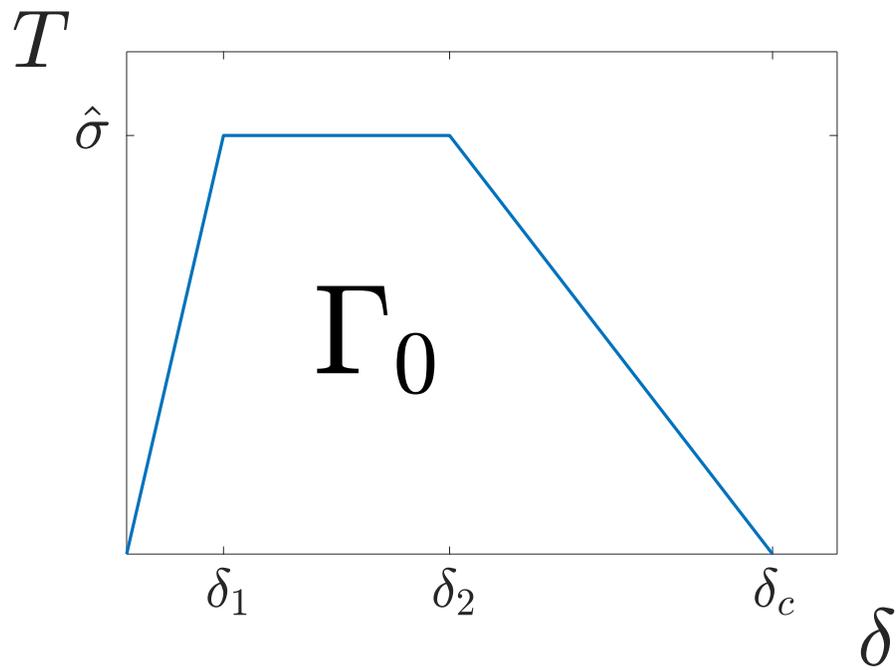}
    \caption{Cohesive traction $T$ - separation $\delta$ law characterising the fracture process.}
    \label{fig:CoheLaw}
\end{figure}

We model tensile fracture at the tip of a mode I crack by means of a cohesive zone model, following Tvergaard and Hutchinson \cite{Tvergaard1992} - see Fig. \ref{fig:Crack}. Cohesive zone formulations have a long history back to Dugdale \cite{Dugdale1960} and Barenblatt \cite{Barenblatt1962}: fracture is regarded as a gradual process in which separation takes place across an extended cohesive zone, and is resisted by cohesive tractions. As shown in Fig. \ref{fig:CoheLaw}, we shall make use of a trapezoidal traction-separation law of strength $\hat{\sigma}$, with its shape being defined by a critical cohesive separation $\delta_c$ and by two shape parameters $\delta_1=0.15 \delta_c$ and $\delta_2=0.5 \delta_c$. The work of fracture $\Gamma_0$ is given by the area under the traction-separation curve, such that
\begin{equation}
\Gamma_0 = \frac{1}{2} \hat{\sigma} \left( \delta_c + \delta_2 - \delta_1 \right)
\end{equation}

A reference stress intensity factor for crack growth initiation follows immediately as
\begin{equation}
K_0 = \sqrt{\frac{E \Gamma_0}{\left(1 - \nu^2 \right)}}
\end{equation}

\noindent along with a reference length $R_0$, where
\begin{equation}
R_0 = \frac{1}{3 \pi \left( 1 - \nu^2 \right)} \frac{E \Gamma_0}{\sigma_0^2} = \frac{1}{3 \pi} \left( \frac{K_0}{\sigma_0} \right)^2
\end{equation}

The crack tip is placed at the origin and the crack plane is aligned with the negative $x_1$ axis of the Cartesian reference frame $(x_1,\, x_2)$. A remote $K_I$ field is imposed by a boundary layer formulation, as follows. The outer periphery of the mesh is subjected to the mode I elastic $K$-field,
\begin{equation}
u_i = \frac{K_I}{E} r^{1/2} f_i \left( \theta, \, \nu \right)
\end{equation}

\noindent where $r$ and $\theta$ are polar coordinates centred at the crack tip and the functions $f_i \left( \theta, \, \nu \right)$ are given by,
\begin{equation}
f_1=\frac{1+\nu}{\sqrt{2 \pi}} \left( 3 - 4 \nu - \cos \theta \right) \cos \left( \frac{\theta}{2} \right)
\end{equation}
\begin{equation}
f_2=\frac{1+\nu}{\sqrt{2 \pi}} \left( 3 - 4 \nu - \cos \theta \right) \sin \left( \frac{\theta}{2} \right)
\end{equation}

\begin{figure}[H] 
    \centering
    \includegraphics[scale=0.7]{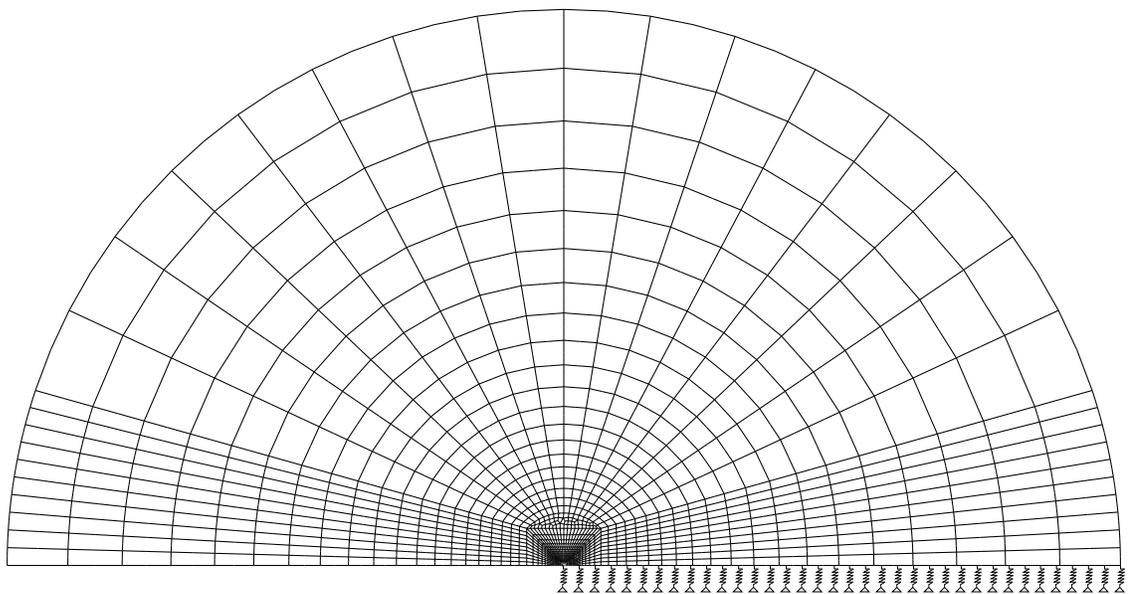}
    \caption{Finite element mesh and configuration of the boundary layer.}
    \label{fig:BL}
\end{figure}

Upon exploiting the symmetry about the crack plane, only half of the model is analysed, as shown in Fig. \ref{fig:BL}. The finite element model is implemented in the commercial finite element package Abaqus \cite{Abaqus2} and we solve the boundary value problem by an implicit Backward Euler integration scheme. Plane strain quadratic quadrilateral elements are employed, with the mesh comprising 267272 degrees of freedom. A refined mesh was used along the cohesive zone in order to obtain a converged solution. The characteristic length of the elements in the crack propagation region is chosen to be equal to $\delta_c$. Cohesive elements with 6 nodes and 12 integration points are implemented by means of a user element (UEL) subroutine \cite{EFM2017}. A control algorithm is used to avoid convergence problems due to snap-back instabilities, see \cite{Segurado2004,TAFM2017} for details. Computations have been performed within an infinitesimal deformation framework since strains remain small, as argued by Tvergaard and Hutchinson \cite{Tvergaard1992} in their finite strain analysis. Dimensional analysis shows that the solution, given in terms of the remote $K_I=K_R (\Delta a)$, is a function $F$ of the following dimensionless quantities,
\begin{equation}
\frac{K_R}{K_0} = F \left( \frac{\Delta a}{R_0}, \, \frac{\sigma_0}{E}, \,  \frac{\hat{\sigma}}{\sigma_0}, \, \frac{E_t}{E}, \, \nu, \, N, \, \frac{\delta_1}{\delta_c}, \, \frac{\delta_2}{\delta_c} \right)
\end{equation}

We will conduct calculations until steady state crack growth at constant $K_{SS}$ is attained. Tvergaard and Hutchinson \cite{Tvergaard1992} showed that $K_{SS}$ is sensitive to the ratio of cohesive strength $\hat{\sigma}$ to material yield strength $\sigma_0$. For an elastic, perfectly plastic solid, $K_{SS}/K_0$ raises steeply as $\hat{\sigma}/\sigma_0$ approaches 3. The interpretation is straightforward by considering a stationary crack in an elastic, ideally plastic solid absent of a cohesive zone. The tensile stress directly ahead of the crack tip equals $3 \sigma_0$ as given by the Prandtl field. Consequently, if $\hat{\sigma}/\sigma_0$ exceeds 3, the crack tip blunts without advance as the cohesive zone strength is not overcome.\\

Tvergaard and Hutchinson \cite{Tvergaard1992} also considered the role of isotropic strain hardening on the R-curve. In this case, the stress field ahead of the crack tip exceeds $3 \sigma_0$ due to the presence of strain hardening. For $\hat{\sigma}/\sigma_0<3$, a shallow R-curve is exhibited and $K_{SS}/K_0$ is slightly above unity. In contrast, for $\hat{\sigma}/\sigma_0>3$, a steeper R-curve is observed and $K_{SS}/K_0$ increases its sensitivity to $\hat{\sigma}$. A major aim of the present study is to explore the sensitivity of the R-curve to the nature of the hardening law: isotropic versus kinematic hardening.

\section{Results}
\label{Sec:Results}

\begin{figure}[H] 
    \centering
    \includegraphics[scale=1.1]{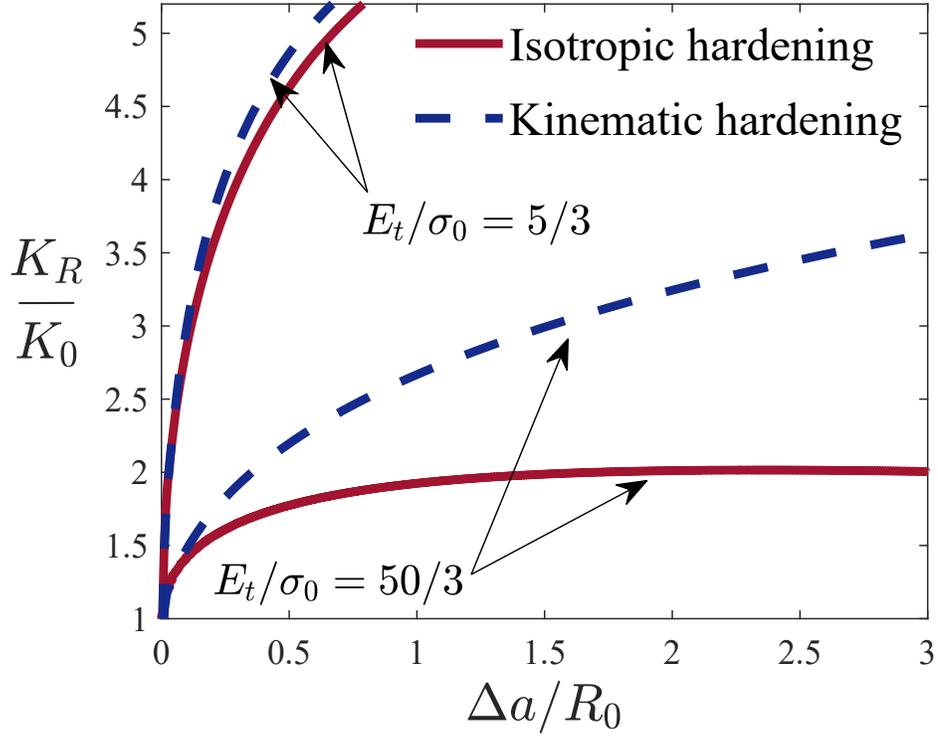}
    \caption{Crack growth resistance curves for linear isotropic and kinematic hardening plasticity and different hardening levels. Material properties: $\delta_1/\delta_c=0.15$, $\delta_2/\delta_c=0.5$, $\sigma_0/E=0.003$, $\nu=0.3$, and $\hat{\sigma}=3.5 \sigma_0$.}
    \label{fig:IsoKinRb}
\end{figure}

R-curves are shown in Fig. \ref{fig:IsoKinRb} for linear hardening and $\hat{\sigma}/\sigma_0=3.5$; this value of $\hat{\sigma}/\sigma_0$ is close to the limiting value of $\hat{\sigma}/\sigma_0=3$ for an elastic, perfectly plastic solid, as discussed in the previous section. Consider first the R-curve for a small level of strain hardening $E_t/\sigma_0=5/3$. A steeply rising R-curve is predicted, which will give rise to a large steady state fracture toughness $K_{SS}/K_0$ and a large value of the crack extension to achieve steady state $\left( \Delta a /R_0 \right)_{SS}$. The steep R-curve is a consequence of plastic dissipation with crack advance. Little difference is observed between the kinematic and isotropic hardening predictions since the degree of hardening is small. We note in passing that for the elastic, ideally plastic case, $E_t / \sigma_0=0$, and $\hat{\sigma}/\sigma_0=3.5$ no crack advanced is observed: the tensile traction ahead of the crack tip is insufficient to overcome the cohesive zone strength. Now consider the case of a high strain hardening rate $E_t/\sigma_0=50/3$. The strain level near the crack tip can now exceed the cohesive strength $\hat{\sigma}$ at a relatively low value of plastic strain. A shallow R-curve is predicted. Again, kinematic hardening elevates the R-curve compared to the isotropic hardening case.\\

\begin{figure}[H] 
    \centering
    \includegraphics[scale=1.1]{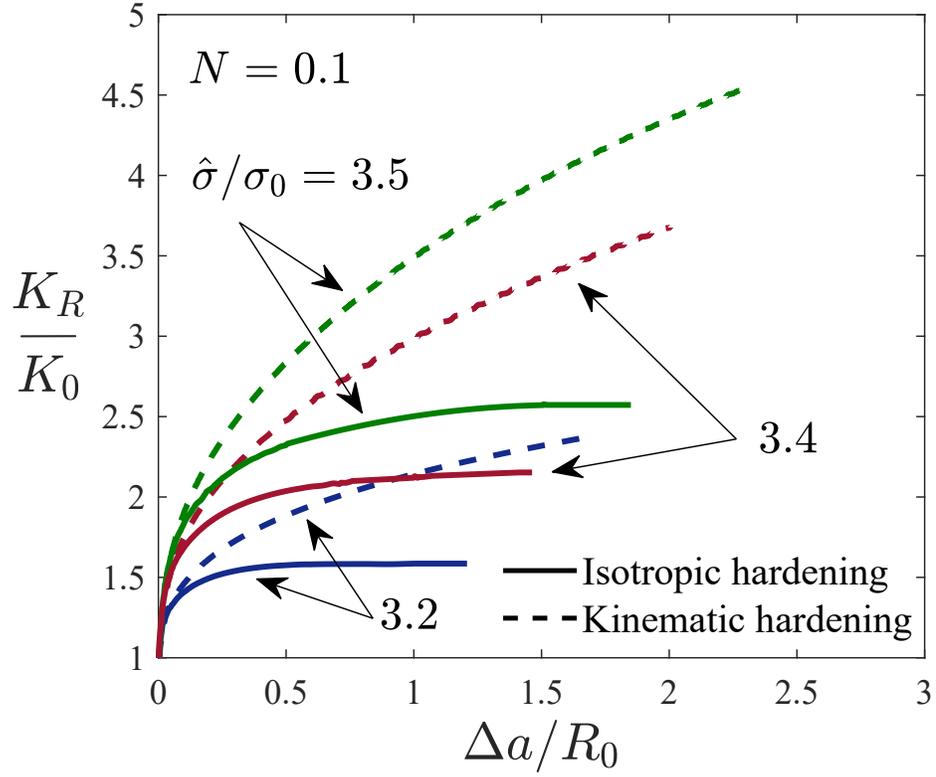}
    \caption{Crack growth resistance curves for power law isotropic and kinematic hardening plasticity and different levels of the cohesive strength. Material properties: $\delta_1/\delta_c=0.15$, $\delta_2/\delta_c=0.5$, $\sigma_0/E=0.003$, $\nu=0.3$, and $N=0.1$.}
    \label{fig:IsoKinN}
\end{figure}

The predicted R-curves for the case of a power law hardening solid are shown in Fig. \ref{fig:IsoKinN} for the choice of $N=0.1$. We consider both isotropic and kinematic hardening, and selected values of the cohesive strength $\hat{\sigma}/\sigma_0=3.2$, $3.4$ and $3.5$. As expected, increasing $\hat{\sigma}/\sigma_0$ elevates the $K$ versus $\Delta a$ response for both hardening laws. However, the R-curves are more sensitive to the cohesive strength for the case of kinematic hardening. Also, as in the linear hardening study, the R-curves are steeper for kinematic hardening, implying a higher value of the steady state fracture toughness $K_{SS}$. We note that there is no straightforward relationship between $K_0$, $K_{SS}$ and the crack initiation toughness $K_{Ic}$ as defined in the standard test methods for fracture toughness, such as the ASTM E 1820 \cite{ASTM1820}. The standard defines $K_{Ic}$ as the value of $K$ corresponding to a crack growth increment which is in the range of 0.2 - 0.5 mm, see \cite{Anderson2005} for a discussion. When the R-curve is steep, $K_{Ic}/K_0$ may be large.\\

\begin{figure}[H] 
    \centering
    \includegraphics[scale=1.1]{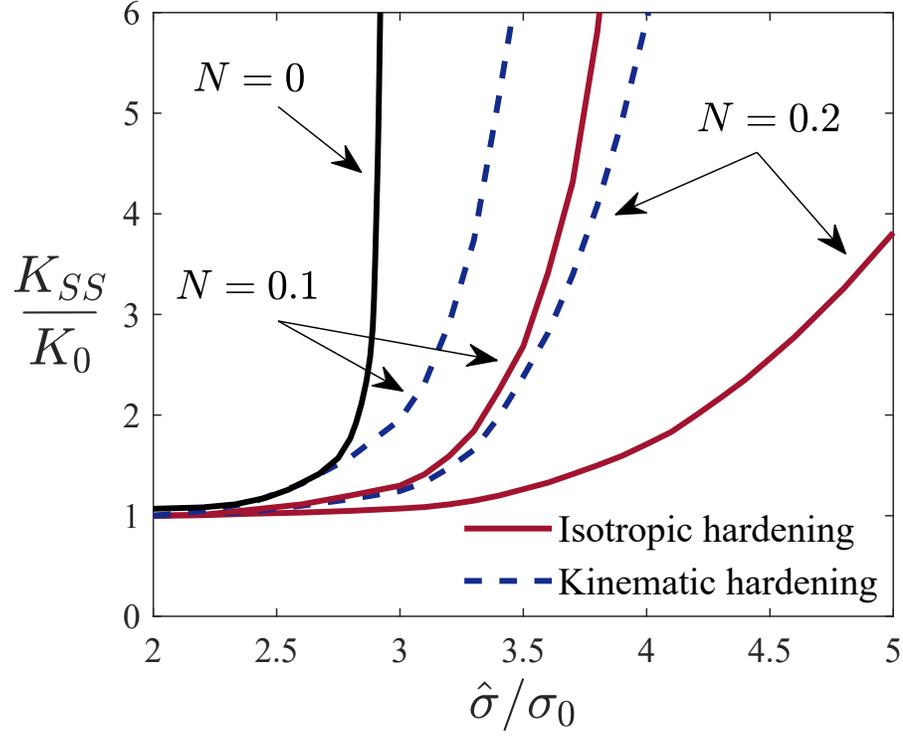}
    \caption{Steady state toughness as a function of the cohesive strength for isotropic and kinematic hardening at different $N$ levels. Material properties: $\delta_1/\delta_c=0.15$, $\delta_2/\delta_c=0.5$, $\sigma_0/E=0.003$, and $\nu=0.3$.}
    \label{fig:SSfigure}
\end{figure}

The steady state toughness is attained when $K_R$ reaches a plateau value. Fig. \ref{fig:SSfigure} shows the sensitivity of $K_{SS}$ to the strain hardening exponent and to the cohesive strength. The isotropic curves are in agreement with the results of Tvergaard and Hutchinson \cite{Tvergaard1992}. First, note that for $N=0$ a single $K_{SS}/K_0$ versus $\hat{\sigma}/\sigma_0$ curve corresponds to the cases of isotropic and kinematic hardening. The value of the steady state toughness increases rapidly in the vicinity of the limiting value of $\hat{\sigma}/\sigma_0=2.8$. With increasing $N$, kinematic and isotropic hardening theories give increasingly divergent predictions. Consistently, for $N>0$, kinematic hardening leads to a higher value of $K_{SS}/K_0$ at a given $\hat{\sigma}/\sigma_0$ than does isotropic hardening. Also, the value of the cohesive strength at which $K_{SS}/K_0$ increases rapidly is lower for the case of kinematic hardening.\\

\begin{figure}[H] 
    \centering
    \includegraphics[scale=1.1]{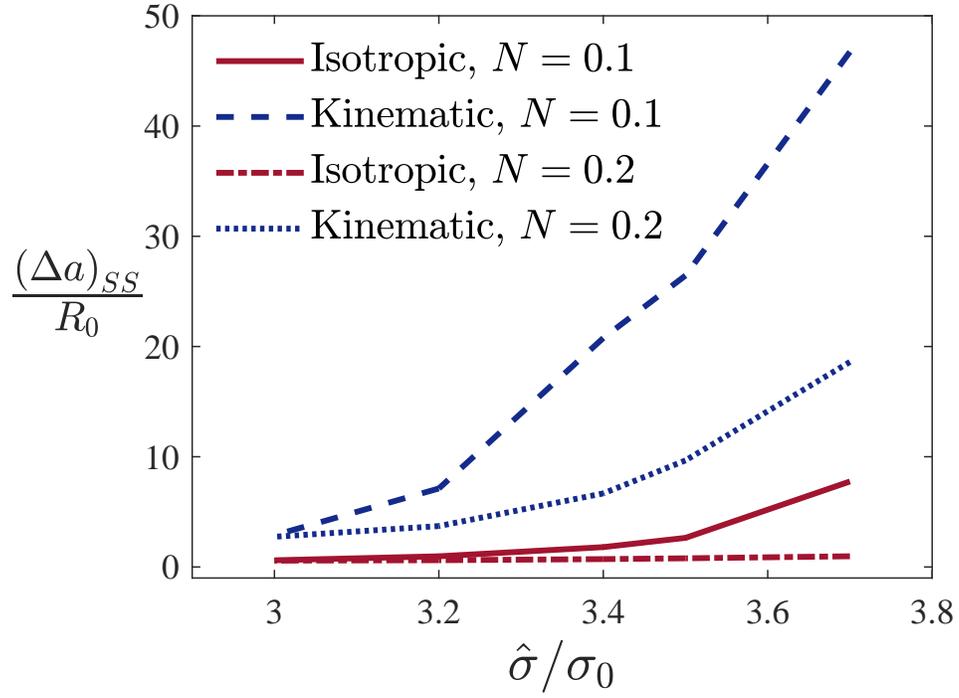}
    \caption{Crack extension at steady state as a function of the cohesive strength for isotropic and kinematic hardening at different $N$ levels. Material properties: $\delta_1/\delta_c=0.15$, $\delta_2/\delta_c=0.5$, $\sigma_0/E=0.003$, and $\nu=0.3$.}
    \label{fig:aSteadyState}
\end{figure}

It is instructive to consider the value of crack extension $\left( \Delta a \right)_{SS}$ that is required to achieve the steady state toughness. The dependence of $\left( \Delta a \right)_{SS}$ upon $\hat{\sigma}/\sigma_0$ is shown in Fig. \ref{fig:aSteadyState} for the power law hardening solid, for both kinematic and isotropic hardening. Note that both $K_{SS}/K_0$ and $\left( \Delta a \right)_{SS}/R_0$ depend upon $\hat{\sigma}/\sigma_0$ in a highly non-linear manner for both hardening laws, recall Figs. \ref{fig:SSfigure} and \ref{fig:aSteadyState}. Is there a simple relation between $\left( \Delta a \right)_{SS}/R_0$ and $K_{SS}/K_0$? This might be expected as the plastic zone size associated with $K=K_{SS}$, is of the order,
\begin{equation}
R_{SS} = \frac{1}{3 \pi} \left( \frac{K_{SS}}{\sigma_0} \right)^2
\end{equation}

Assume that $\left( \Delta a \right)_{SS}$ is proportional to $R_{SS}$,
\begin{equation}
\left( \Delta a \right)_{SS} = C R_{SS}
\end{equation}

\noindent where the constant $C$ is of order unity with some sensitivity to the choice of the hardening law and to $N$. It follows immediately that,
\begin{equation}\label{Eq:aSS}
\frac{\left( \Delta a \right)_{SS}}{R_0} = C \frac{R_{SS}}{R_0} = C \left(\frac{K_{SS}}{K_0} \right)^2
\end{equation}

\begin{figure}[H] 
    \centering
    \includegraphics[scale=1.1]{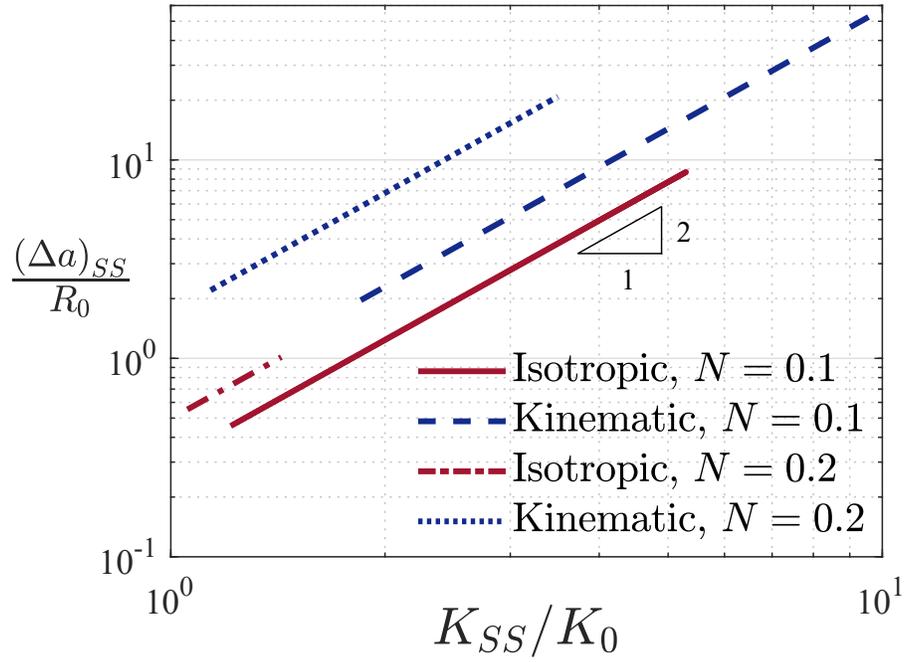}
    \caption{Log-log representation of the relation between the crack extension at steady state and the steady state fracture toughness for isotropic and kinematic hardening at different $N$ levels. Each data point corresponds to a value of the cohesive strength. Material properties: $\delta_1/\delta_c=0.15$, $\delta_2/\delta_c=0.5$, $\sigma_0/E=0.003$, and $\nu=0.3$.}
    \label{fig:aSSvsKsslog}
\end{figure}

The accuracy of this prediction is shown by a cross-plot of $\left( \Delta a \right)_{SS}/R_0$ versus $K_{SS}/K_0$ in Fig. \ref{fig:aSSvsKsslog}, with $\hat{\sigma}/\sigma_0$ as the parameter trending variable. A curve fit reveals that $C$ increases from 0.574 to 1.703 as $N$ goes from 0.1 to 0.2 for kinematic hardening, and $C$ increases from 0.31 to 0.496 as $N$ goes from 0.1 to 0.2 for isotropic hardening. Our numerical predictions show that the isotropic hardening idealization may significantly underestimate the degree of subcritical crack propagation before catastrophic failure. Note further that (\ref{Eq:aSS}) can be re-expressed in the form
\begin{equation}
\Delta a_{SS} = \frac{C}{3 \pi} \left( \frac{K_{SS}}{\sigma_0} \right)^2
\end{equation}

\begin{figure}[H] 
    \centering
    \includegraphics[scale=0.91]{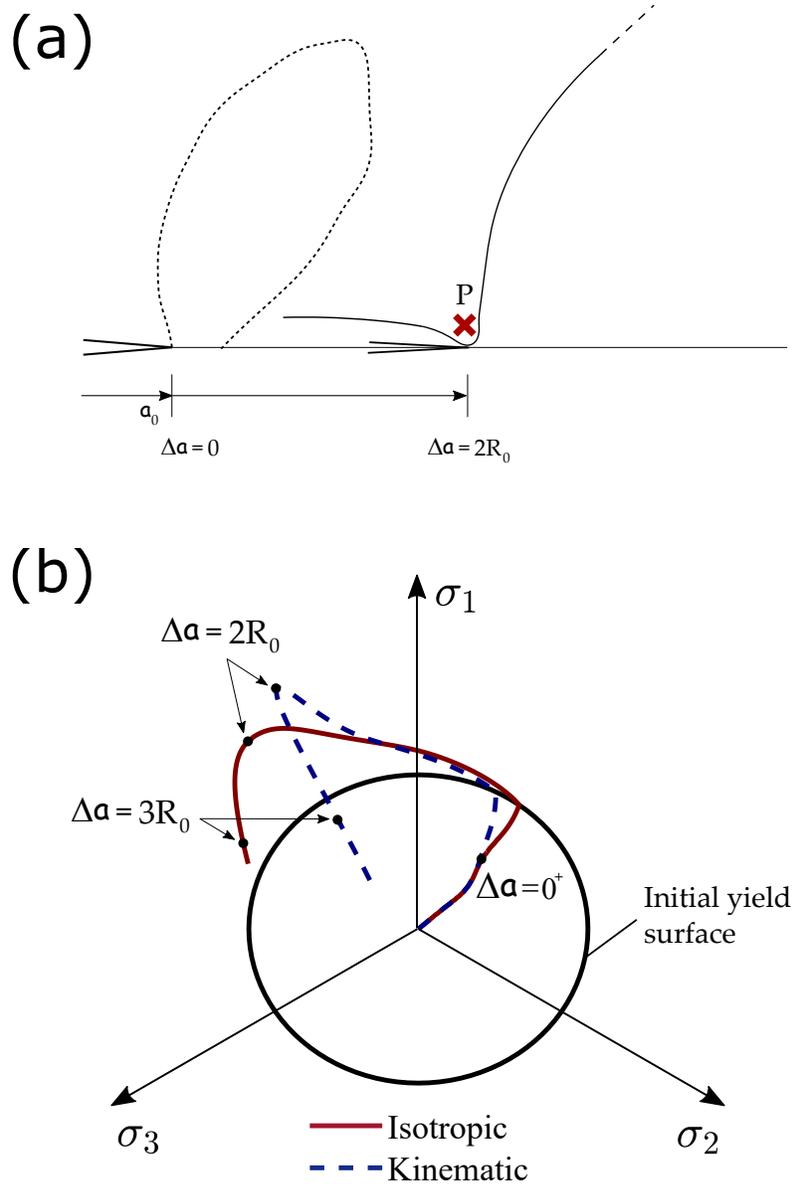}
    \caption{Schematic insight into the effect of isotropic or kinematic hardening on a material point ahead of the initial crack ($r=2R_0$); (a) active plastic zone and evolution path, and (b) stress state on the $\pi$-plane. Material properties: $\delta_1/\delta_c=0.15$, $\delta_2/\delta_c=0.5$, $\sigma_0/E=0.003$, $\nu=0.3$, $\hat{\sigma}/\sigma_Y=3.7$, and $N=0.1$.}
    \label{fig:PiPlanePaths}
\end{figure}

What is the physical basis for the steeper R-curve observed in the case of kinematic hardening? We show in Fig. \ref{fig:PiPlanePaths} that significant non-proportional loading occurs in the vicinity of the crack tip, as the crack advances. Consider a representative material point $P$ at a distance for $2R_0$ ahead of the initial crack tip and slightly above the cracking plane (height of $0.1R_0$). Allow the crack to advance by $\Delta a = 2 R_0$ for both cases of isotropic and kinematic hardening. The active plastic zone is shown in Fig. \ref{fig:PiPlanePaths}a for $\Delta a=0^+$ and $\Delta a = 2 R_0$. The plastic zone at $\Delta a=0^+$ for kinematic hardening is identical to that for isotropic hardening, whereas the plastic zone at $\Delta a = 2 R_0$ is much larger in the kinematic hardening case. Only the isotropic hardening active plastic zone is shown at $\Delta a = 2 R_0$ for the sake of clarity. The stress paths imposed on point $P$ for isotropic and kinematic hardening are given in Fig. \ref{fig:PiPlanePaths}b. Differences between kinematic and isotropic stress paths arise soon after cracking initiates ($\Delta a=0^+$), due to non-proportional straining in the neighbouring points. As the crack advances not only are the stress paths non-proportional but they also deviate from each other, with the greatest change in stress direction given by isotropic hardening. The stronger path dependence of kinematic hardening also plays an important role on localization in thin sheets and in shear localization. For example, Tvergaard \cite{Tvergaard1978} showed that the forming limit curves predicted by kinematic hardening are in better agreement with experimental results than isotropic hardening predictions. The dependence of the critical strain for shear localization upon the local curvature of the yield surface has been investigated by Mear and Hutchinson \cite{Mear1985}.\\

In addition, we investigate the level of energy dissipation in the main plastic zone in the vicinity of the crack tip (denoted $W_1$) and in the secondary plastic zone that arises in the crack wake (denoted as $W_2$). Here, the energy dissipated due to plastic deformation is computed for all material elements in the \emph{active} plastic zone as,
\begin{equation}
W \left( \Delta a \right) = \int_{0}^{\Delta a} \left( \int \sigma_{ij} \dot{\varepsilon}^p_{ij} \textnormal{dV} \right) \textnormal{d}a
\end{equation}

The predictions are given in Table \ref{Tab:PlasticDissipation} for various stages of crack advance. Computations reveal that plastic work in the secondary plastic region is negligible relative to the energy dissipated in the vicinity of the crack. We conclude that differences between isotropic and kinematic hardening responses are mainly due to non-proportional deformation in the crack tip plastic zone. Also, Table \ref{Tab:PlasticDissipation} shows that kinematic hardening involves a much larger plastic dissipation energy than isotropic hardening, and is consistent with the steeper R-curve.\\

\begin{table}[H]
\centering
\caption{Plastic energy dissipation with crack advance in the primary plastic zone region at the crack tip $W_1$ and the secondary plastic zone region at the crack wake $W_2$. Material properties: $\delta_1/\delta_c=0.15$, $\delta_2/\delta_c=0.5$, $\sigma_0/E=0.003$, $\nu=0.3$, $N=0.1$ and $\hat{\sigma}=3.5 \sigma_0$.}
\label{Tab:PlasticDissipation}
   {\tabulinesep=1.2mm
   \begin{tabu} {|c|c|c|c|c|}
       \hline
  \multirow{2}{*}{$\Delta a/ R_0$} & \multicolumn{2}{ |c| }{Isotropic} & \multicolumn{2}{ |c| }{Kinematic} \\ \cline{2-5}
   & $W_1/\left( \Gamma_0 \Delta a \right)$ & $W_2/\left( \Gamma_0 \Delta a \right)$ &  $W_1/\left( \Gamma_0 \Delta a \right)$ & $W_2/\left( \Gamma_0 \Delta a \right)$\\ \hline
 0.5 & 21.98 & 0.01 & 52.69 & 0.42 \\
 1 & 17.24 & 0.14 & 58.36 & 0.85 \\
 2 & 8.21 & 0.17 & 78.75 & 1.08\\\hline
   \end{tabu}}
\end{table}

\section{Conclusions}
\label{Sec:Concluding remarks}

We investigated how the isotropic or kinematic nature of strain hardening influences crack growth resistance. Finite element results show very significant differences between isotropic and kinematic hardening laws that yield the same response under uniaxial tension. We show that kinematic hardening notably enhances plastic dissipation and the steady state fracture toughness $K_{SS}$. These differences persist over different hardening levels, cohesive strengths, and hardening profiles. 

\section{Acknowledgments}
\label{Acknowledge of funding}

The authors acknowledge valuable discussions with V.S. Deshpande (University of Cambridge). E. Mart\'{\i}nez-Pa\~neda also acknowledges valuable insight from K. Juul and C.F. Niordson (Technical University of Denmark). The authors would like to acknowledge the funding and technical support from BP (ICAM02ex) through the BP International Centre for Advanced Materials (BP-ICAM). E. Mart\'{\i}nez-Pa\~neda additionally acknowledges financial support from the Ministry of Economy and Competitiveness of Spain through grant MAT2014-58738-C3 and the People Programme (Marie Curie Actions) of the European Union's Seventh Framework Programme (FP7/2007-2013) under REA grant agreement n$^{\circ}$ 609405 (COFUNDPostdocDTU). 




\bibliographystyle{elsarticle-num}
\bibliography{library}

\end{document}